\documentclass[<review>]{elsarticle}
\usepackage{amssymb, amsmath, mathrsfs, epsfig, titlesec}

\makeatletter
\def\ps@pprintTitle{%
 \let\@oddhead\@empty
 \let\@evenhead\@empty
 \def\@oddfoot{\centerline{\thepage}}%
 \let\@evenfoot\@oddfoot}
\makeatother

\begin{document}
\begin{frontmatter}

\title{MATHEMATICAL MODELING OF PERIFUSION CELL CULTURE EXPERIMENTS ON GNRH SIGNALING }

\author[dukemath]{N Ezgi Temamogullari \corref{cor}}
\ead{ezgi@math.duke.edu}

\author[dukebio]{H Frederik Nijhout}
\ead{hfn@duke.edu}

\author[dukemath]{Michael C Reed}
\ead{reed@math.duke.edu}

\cortext[cor]{Corresponding author}
\address[dukemath]{Department of Mathematics, Duke University, Durham, North Carolina}
\address[dukebio]{Department of Biology, Duke University, Durham, North Carolina}

\begin{abstract}
The effects of pulsatile GnRH stimulation on anterior pituitary cells are studied using perifusion cell cultures, where constantly moving medium over the immobilized cells allows intermittent GnRH delivery. The LH content of the outgoing medium serves as a readout of the GnRH signaling pathway activation in the cells. The challenge lies in relating the LH content of the medium leaving the chamber to the cellular processes producing LH secretion. To investigate this relation we developed and analyzed a mathematical model consisting of coupled partial differential equations describing LH secretion in a perifusion cell culture. We match the mathematical model to three different data sets and give cellular mechanisms that explain the data. Our model illustrates the importance of the negative feedback in the signaling pathway and receptor desensitization. We demonstrate that different LH outcomes in oxytocin and GnRH stimulations might originate from different receptor dynamics and concentration. We analyze the model to understand the influence of parameters, like the velocity of the medium flow or the fraction collection time, on the LH outcomes. We show that slow velocities lead to high LH outcomes. Also, we show that fraction collection times, which do not divide the GnRH pulse period evenly, lead to irregularities in the data. We examine the influence of the rate of binding and dissociation of GnRH on the GnRH movement down the chamber. Our model serves as an important tool that can help in the design of perifusion experiments and the interpretation of results.  
\end{abstract}

\begin{keyword}
Mathematical Model \sep Perifusion Cell Culture \sep GnRH  \sep LH 
\end{keyword}
\end{frontmatter}

\newpage
\section{INTRODUCTION} 
\begin{figure}[b] \centering 
\includegraphics[width=\textwidth]{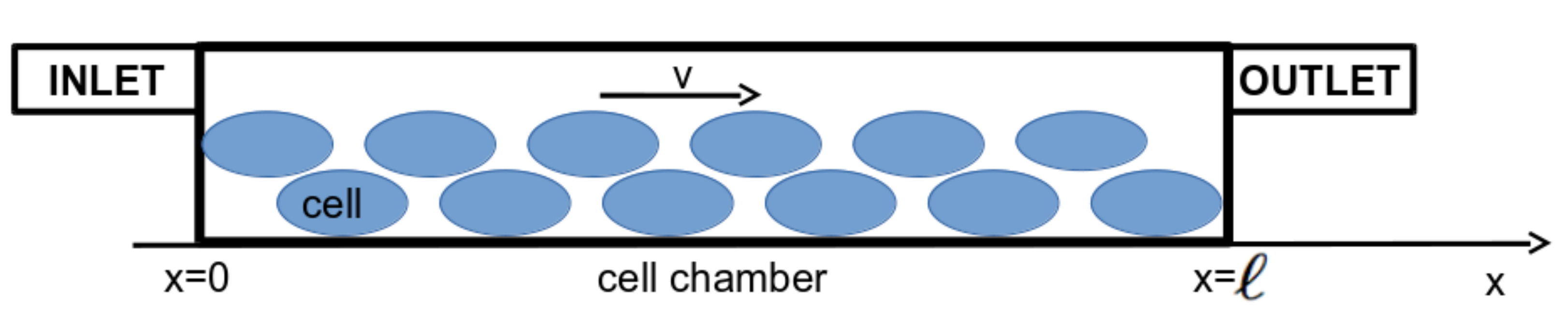} 
\caption{ {\footnotesize{\textbf{Perifusion Cell Culture Diagram.} The blue circles are the cells immobilized in the cell chamber. The culture medium enters the chamber through the inlet, flows with velocity \textit{v} through the chamber and leaves the chamber through the outlet. The length of the cell chamber is $\ell$.}} }
\label{fig:perifusionchamber}
\end{figure}

\begin{figure}[t] \centering 
\includegraphics[width=\textwidth]{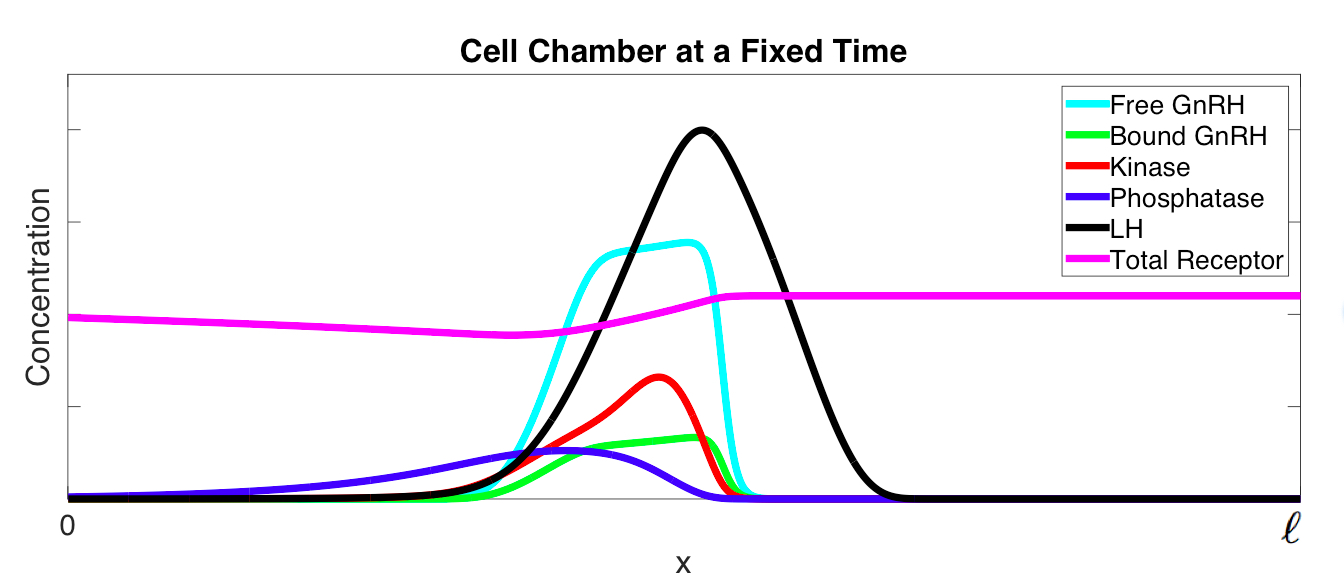} 
\caption{ {\footnotesize{\textbf{Simulation of the Perifusion Cell Chamber at a Fixed Time}. The x-axis is the position in the cell chamber from 0 to $\ell$; the y-axis is the concentration of the species in arbitrary units.}}}
\label{fig:chamberfixedtime}
\end{figure}

Gonadotropin Releasing Hormone (GnRH) is the master regulator of reproductive physiology. It is secreted from the hypothalamus in pulses and stimulates the anterior pituitary gonadotroph cells via GnRH receptors (GnRH-R). This stimulation leads to synthesis and secretion of the gonadotropins Luteinizing Hormone (LH) and Follicle Stimulating Hormone (FSH), which regulate sex steroid hormone synthesis and progression through the menstrual cycle. \\

The influence of the pulsatile GnRH signal on pituitary cells can be studied  \textit{in vitro} with perifusion cell cultures \cite{weiss1993perifusion, bedecarrats2003differential, purwana2010induction, garrel2010sustained,  pemberton2014pi3k, haisenleder2008pulsatile, evans2001release, mcintosh1985dynamic}, in which the cells are immobilized in a cell chamber, through which the culture medium flows continuously at a constant rate (see Figure \ref{fig:perifusionchamber}). GnRH is introduced into the cell chamber along with the fresh medium through the inlet, where the concentration of GnRH entering the chamber per unit time is controlled. The effluent is collected in fixed time intervals and these fractions are analyzed for their LH content. GnRH moves down the chamber with the moving fluid and as it travels, it binds to GnRH-Rs on the cells activating a signaling cascade. The activation of the signaling cascade causes LH release from the cells into the cell chamber and the secreted LH travels with the flowing medium towards the outlet.\\

The GnRH signaling pathway in gonadotrophs involves activation of different mitogen-activated protein kinases (MAPK) and MAPK phosphatases (MKP) \cite{naor2000activation,ruf2004genomics,  perrett2013molecular, thackray2010hormones}. MAPKs become activated by phosphorylation and active MAPKs induce gonadotropin synthesis and secretion. Active MKPs inactivate MAPKs by dephosphorylating them, and thus, MKPs constitute an important negative feedback mechanism \cite{zhang2006role, ciccone2009biology, kanasaki2011extracellular,thompson2014gnrh}. GnRH-R desensitization constitutes another negative feedback mechanism \cite{mcardle2002signalling, dobkin2006activation}. All of these mechanisms are part of our mathematical model. The GnRH signaling pathway has other downstream components that we do not consider. \\

The interesting mathematical question is how to use the effluent LH content to understand the cellular processes producing LH. Diverse mathematical models of the GnRH signaling pathway have been developed for analyzing perifusion cell culture data \cite{smith1991mathematical, magill2013mathematical, blum2000mathematical, washington2004theoretical, evans2013two, bertram2008mathematical, heinze1998mathematical}. However, to the best of our knowledge, almost all of them are restricted to systems of ordinary differential equations lacking the spatial component of the experimental setup. As we will see, spatial parameters such as the length of the chamber and the velocity of the medium have a significant effect in the experimental results. \cite{smith1991mathematical} analyzes partial differential equations describing how a signal distorts as it passes through a chamber, but it lacks the signal transduction in the cells. \\

In this paper we present a novel mathematical model of the LH secretion from gonadotroph cells in a perifusion cell culture. The model consists of a system of coupled partial differential equations describing the movement of GnRH and LH in the cell chamber and the dynamics of total receptors, kinases and phosphatases in the cells. Figure \ref{fig:chamberfixedtime} shows a simulation depicting the cell chamber at a fixed time. The x-axis shows the position in the chamber and the y-axis shows the concentrations of free GnRH (cyan), bound GnRH (green), total receptors (magenta), active kinase (red), active phosphatase (blue) and LH (black). As the free GnRH (cyan) moves down the chamber, it binds to the GnRH-Rs forming bound GnRH (green) and activates the kinases (red). This leads to both activation of phosphatases (blue) and the release of LH (black) into the chamber. Released LH moves down the chamber with the medium. In this simulation LH moves faster than the free and bound GnRH, since the movement of GnRH is retarded by its interaction with the receptors. Bound GnRH, in addition to activating the signaling cascade, leads to a decrease in the total receptor concentration (magenta) via desensitization of the receptor. The mathematical model presented in this paper aids in understanding and interpreting perifusion cell culture data by connecting the dynamics of the LH outcome to the cellular processes generating it. The model also shows how the parameters controlled by the experimenters, like the velocity of the fluid flow, the cell concentration in the chamber, the GnRH pulse characteristics, affect the experimental outcomes.\\

In Section \ref{mathematicalmodel} we describe the mathematical model. In Section \ref{perifusion} we compare the mathematical model to three different data sets consisting of the LH content of the fractions collected from the cell chamber. The first data set, taken from \cite{cantor1996use}, is a finely sampled data, where LH shows a characteristic triphasic response. We use the first data set to estimate the model parameters and we explain which cellular interactions might be underlying the shape of the LH output. The second data set \cite{mcintosh1986varying} is collected over a longer time course, where multiple GnRH pulses are introduced into the chamber. The same parameters used in the first data set reproduce the second data set, thus, the second data set serves as a cross validation. Also, using the second data set we show how the length of the fraction collection time affects the regularity of the data. The third data set \cite{gonzalez2014direct} compares oxytocin and GnRH action on pituitary cells. We use a different set of parameters for the third data set, which is biologically reasonable, since the cell type used is different than the first two data sets. We matched the parameters to GnRH stimulation data and recovered the oxytocin stimulation outcome just by changing one or two parameters related to the receptor dynamics, showing a possible mechanism explaining the different outcomes in these two cases. In Section \ref{insilico}, we conduct \textit{in silico} experiments to investigate the importance of key parameters. First, we show how the velocity of the medium flow affects the LH outcome. Next, we focus on the GnRH movement in the chamber and describe how this movement depends on the binding rate of GnRH to its receptor, the dissociation rate of bound GnRH from the receptor and the total GnRH-R concentration. Finally, we discuss how the LH outcome per GnRH amount supplied to the chamber depends on the GnRH pulse characteristics. \\ 

\section{MATHEMATICAL MODEL} \label{mathematicalmodel}
The mathematical model is a system of partial differential equations for the following variables: the free GnRH, $F(x,t)$; GnRH bound to its receptor, $B(x,t)$; total receptor concentration, $R(x,t)$; active kinase, $K(x,t)$; active phosphatase, $P(x,t)$; and the secreted product $L(x,t)$. In the rest of the paper, kinase and phosphatase will refer to the active kinase and the active phosphatase. We assume that the cell chamber is homogeneous at the cross section, thus, we only considered one space dimension, $x$, the distance from the left end point of the chamber. As indicated in Figure \ref{fig:perifusionchamber}, $x$ varies from 0 to $\ell$, where 0 corresponds to the left end of the perifusion chamber and $\ell$ is the length of the chamber.\\

The system of partial differential equations on the domain $x \in [0, \ell]$ and $t \geq 0$ is:\\
\begin{subequations} \label{eq:sysPDE}
\begin{align} 
F_t(x,t)+vF_x(x,t) &=-k_1(R(x,t)-B(x,t))F(x,t)+k_2B(x,t) \label{eq:F} \\
B_t(x,t) &=k_1(R(x,t)-B(x,t))F(x,t)-k_2B(x,t) \label{eq:B} \\
R_t(x,t)&=a_0-b_0R(x,t)-c_0B(x,t) \label{eq:R} \\
K_t(x,t) &=b_1B(x,t)-a_1K(x,t)P(x,t) \label{eq:K} \\
P_t(x,t)&=b_2K(x,t)-a_2P(x,t)  \label{eq:P} \\
L_t(x,t)+vL_x(x,t)&=bs+b_3K(x,t) \label{eq:L}
\end{align}
\end{subequations}
with initial conditions: 
\begin{subequations} \label{eq:IC}
\begin{align}
F(x,0)&=B(x,0)=K(x,0)=P(x,0)=L(x,0)=0 \\ 
R(x,0)&=R_{in}, \text{ for } 0<x\leq \ell 
\end{align}
\end{subequations}
and boundary conditions:
\begin{subequations}
\begin{align} 
B(0,t) &=K(0,t)=R(0,t)=P(0,t)=L(0,t)=0 \label{eq:BC} \\
F(0,t) &= f(x,t) \label{eq:BCF}
\end{align}
\end{subequations}

The meaning of parameters is given in Table \ref{table:parameters}. Equation \eqref{eq:F} describes the evolution of the free GnRH. The advection term $vF_x(x,t)$ models the movement of the free GnRH down the chamber. The right hand sides of the equations \eqref{eq:F} and \eqref{eq:B} describe the binding and dissociation of GnRH to and from its receptor. The term $R(x,t)-B(x,t)$ gives the concentration of the free receptors to which free GnRH can bind. Unlike \eqref{eq:F}, the equation \eqref{eq:B} does not have an advection term, since, once GnRH is bound to its receptor, it cannot move. Similarly, the equations for the total receptors \eqref{eq:R}, the kinase \eqref{eq:K} and the phosphatase \eqref{eq:P} do not have the advection term, since they are associated with the cells and their positions in the cell chamber do not change. The receptors are produced with rate $a_0$ and are removed from the cell membrane with rate $b_0$, which describe the GnRH independent receptor dynamics. Bound GnRH leads to desensitization of the GnRH receptors, as described by the term $-c_0B(x,t)$ in the equation \eqref{eq:R}, forming a negative feedback loop. The kinases are activated by the bound GnRH and inactivated by phosphatases represented by the $b_1B(x,t)$ and $-a_1K(x,t)P(x,t)$ terms in the equation \eqref{eq:K}. The phosphatases are activated by kinases, and they are deactivated with rate $a_{2}$ as given in the equation \eqref{eq:P}. The equation describing the concentration of the LH \eqref{eq:L} has an advection term, $vL_x(x,t)$, because secreted LH moves down the cell chamber with the fluid. LH has two source terms: $bs$ models the GnRH independent basal LH secretion and the $b_3K(x,t)$ term models GnRH dependent secretion, where activation of the kinases lead to LH secretion. We assume that the secretion is instantaneous. \\ 

The parameter $bs$ can be calculated from the GnRH independent steady state basal LH level at $x=\ell$. When the steady state is achieved, $L_t(x,t)$ term will be zero in the equation \eqref{eq:L}. Since the basal LH secretion is independent of GnRH stimulation, $K(x,t)$ term will be zero too. So the steady state basal LH will satisfy $vL_x(x,t)=bs$. At $x=\ell$, the steady state basal LH is equal to $(bs \times \ell)/v$. \\

Initially there are no GnRH, kinase, phosphatase or LH in the chamber and the initial total receptor concentration, $R_{in}$, is assumed to be the same at every point, as given by the initial conditions \eqref{eq:IC}. We assume that there is no cell at the point $x=0$, thus for all times there is no bound GnRH, no GnRH receptor, kinase, phosphatase or LH at $x=0$, as given by the boundary conditions \eqref{eq:BC}. We model the free GnRH input into the chamber as a boundary condition. If $A$ nM of GnRH is introduced into the chamber for $\tau$ minutes, than in the equation \eqref{eq:BCF}, $f(x,t)=A$ for $0 \leq t \leq \tau$ and $f(x,t)=0$ otherwise. In perifusion experiments the GnRH is given as a single pulse or as a train of pulses. To model the latter case, we incorporate the period of the pulses into the boundary condition. \\

We ignore diffusion, since we assume that the advection is the predominant way of mass transport in the systems discussed in this paper based on the following argument: The \textit{Peclet} (Pe) \textit{number}, $Pe= (v \ell)/D$, gives the ratio of mass transport through advection and through diffusion \cite{fournier2011basic}. $v$ is the velocity of the fluid flow, $\ell$ is the length along which the mass transfer happens and $D$ is the diffusion coefficient of the transported substance. The Pe number of the perifusion system discussed in Section \ref{midgley} is approximately $3700$ for GnRH and $19000$ for LH, because the velocity is 0.2 mm/sec, $\ell$ is 5.56 mm, the diffusion coefficient of GnRH is 3.04$\times 10^{-6}$ cm$^2$/sec \cite{heinrich1985effect} and the diffusion coefficient of LH is $6 \times 10^{-7}$ cm$^2$/sec \cite{li1952physicochemical}. We assume that LH and FSH have similar diffusion coefficients, since they have similar molecular weights. Since the Pe numbers are $>>1$ for the transported substances, advection dominates diffusion in this system. \\

\begin{table}[]
\caption{Parameters		2 in the Model}
\begin{center}
\begin{tabular}{ l l l l}
\hline 
$v$ & velocity of the medium flow & $a_1$ & deactivation rate of kinase \\
$\ell$ & length of the cell chamber & $b_2$ & activation rate of phosphatase \\
$k_1$ & binding rate of free GnRH & $a_2$ & deactivation rate of phosphatase \\ 
$k_2$ & dissociation rate of bound GnRH & $bs$ & basal LH secretion rate \\
$a_0$ & synthesis rate of GnRH-R & $R_{in}$ & initial GnRH-R concentration \\ 
$b_0$ & internalization rate of GnRH-R & $b_3$ & LH secretion rate \\
$c_0$ & desensitization rate of GnRH-R & $A$ & delivered free GnRH concentration \\
$b_1$ & activation rate of kinase & period & period of pulses \\
$\tau$ & pulse duration\\ 
\hline
\end{tabular}
\end{center}
\label{table:parameters} 
\end{table}
\newpage
\section{RESULTS}
\subsection{Perifusion Experiments} \label{perifusion}
\subsubsection{Triphasic Response} \label{midgley} Cantor et al. \cite{cantor1996use} developed a microperifusion system for monitoring LH release from gonadotrophs using adult female sheep anterior pituitary fragments . The medium coming from the cell chamber was collected at 30-second intervals and the LH content of these fractions was determined. The cells in the perifusion chamber secreted a basal level of LH before the introduction of GnRH. The GnRH dependent LH response at $x=\ell$ had three phases. Upon introduction of GnRH there was a rapid response forming an initial peak, which was followed by a lower steady state level of LH during the GnRH stimulation. Finally, upon cessation of the GnRH stimulation LH slowly returned to the basal level.\\

\begin{figure} \centering 
\includegraphics[width=\textwidth, scale=0.3]{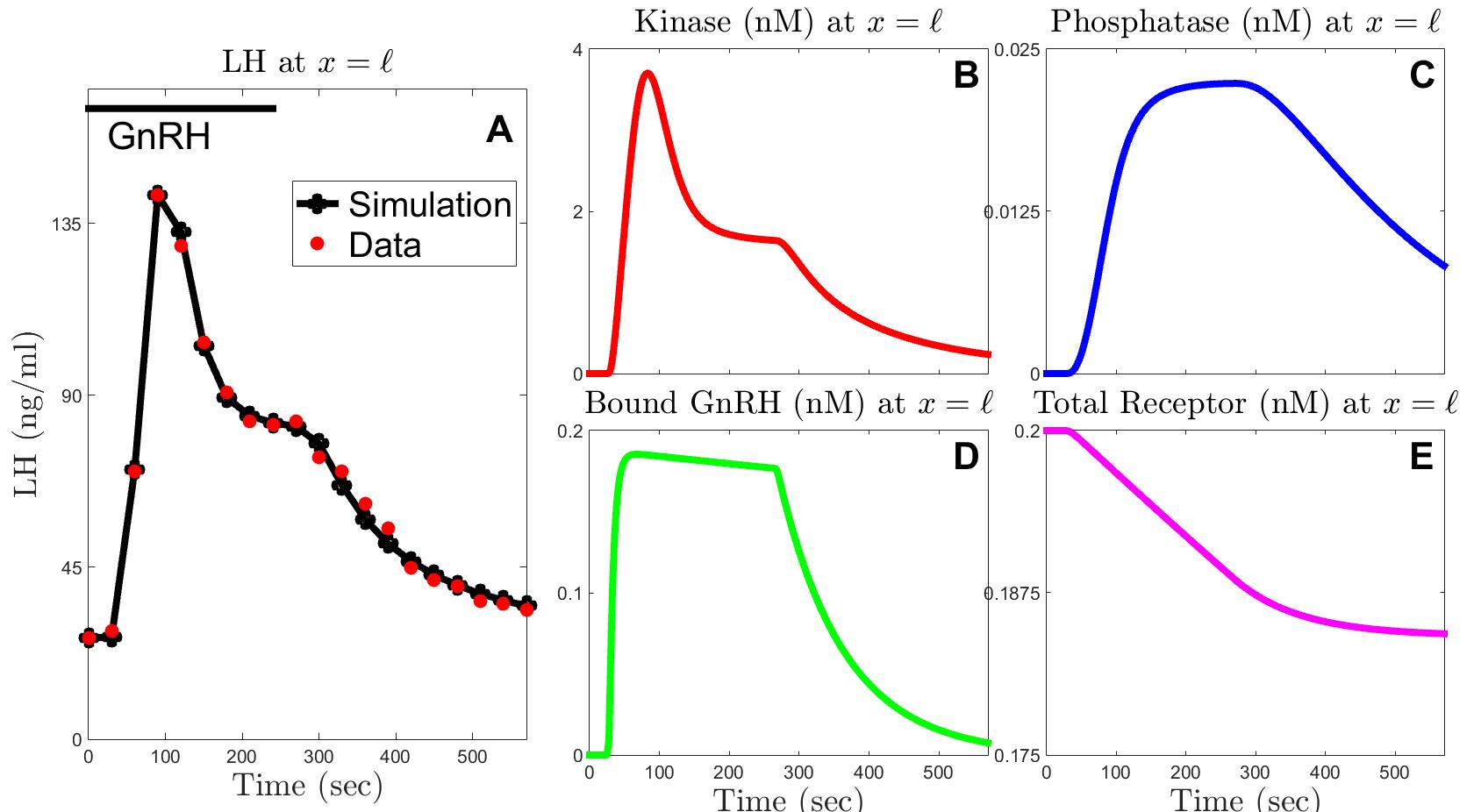} 
\caption{{\footnotesize{\textbf{Time Course of Species at $x=\ell$.} Panel A shows the amount of LH collected in the 30-second fractions in ng/ml. The red dots are the data points, whereas the black dots and the black line connecting them are the simulation results. The 240 second GnRH stimulation is indicated at the top of the graph with the black line. Panels B, C, D and E show the simulation result for the time courses of kinase, phosphatase, bound GnRH and total receptor concentrations at $x=\ell$. All simulations last 570 seconds. Note that the scales are different in each panel.}}}
\label{fig:midgley}
\end{figure}

\textbf{Model Simulation}, The LH concentrations of 30-second collections are shown in Figure \ref{fig:midgley}, Panel A, where the red dots are the data points taken from \cite{cantor1996use}. The black points and the black line collecting them are the simulation results. The time required for the simulation to reach the steady state basal LH level is not shown. The GnRH input, which is represented by the black line at the top, is given to the chamber for 240 seconds. Other panels in Figure \ref{fig:midgley} show simulation results for the time courses of kinase, phosphatase, bound GnRH and total receptor concentrations at $x=\ell$ for 570 sec. When interpreting these graphs, keep in mind that the LH profile in Panel A is the result of the collective action of all kinases at all locations in the chamber.\\

In Panel A, as seen in the first two red dots, the LH content in the first two collections are equal, since the LH secreted at the left end of the cell chamber takes time to travel down to the right end of the chamber. More specifically, since the length of the column is 5.56 mm and the velocity of the medium flow is 0.2 mm/sec, the passage time for LH is 27.8 seconds. Similarly, in the other panels in Figure \ref{fig:midgley}, the concentrations at $x=\ell$ do not rise in the first approximately 30 seconds until the free GnRH reaches $x=\ell$ and initiates the activation of the signaling cascade at that point. With the parameters used in this simulation free GnRH also moves with velocity 0.2 mm/sec, however with different parameters it can have a different velocity. For a detailed discussion of the movement of GnRH see section \ref{GnRHmovement}. After this initial flat line in the LH content of the first two collections, we see a rapid increase in the LH content of fractions, which constitutes the first part of the characteristic triphasic response. Likewise, kinase (Panel B), phosphatase (Panel C) and bound GnRH (Panel D) concentrations rapidly increase following the initial flat line. This rapid initial increase of kinases in the chamber leads to the initial rise in LH secretion.\\

After achieving its peak value, the LH content of the collections decreases to a lower plateau and remains steady until about 30 seconds after the end of the GnRH introduction into the chamber. This approximate 30 seconds delay is the time required for the last part of the GnRH signal introduced at the left end point of the chamber to reach the right end point. This steady LH level originate from the quasi steady state levels reached by the kinase, phosphatase and bound GnRH in this time period. The steady level of the bound GnRH decreases slightly, as the total receptor concentration at $x=\ell$ also slightly decreases due to the receptor desensitization.\\

After this quasi steady-state level, the LH concentration decreases gradually back to the basal LH secretion level (Panel A), forming the last phase of the triphasic response. Upon cessation of GnRH introduction into the chamber, the bound GnRH levels start to drop. Bound GnRH at $x=\ell$ decreases after about 270 seconds, where 240 seconds is the length of GnHR stimulation and 30 seconds is approximately the passage time through the chamber. The decrease in bound GnRH leads to a decrease in the kinase (Panel B) and in turn in the LH levels (Panel A). The rate by which LH returns to the basal level is primarily determined by the dissociation constant of bound GnRH, $k_2$.\\

The total receptor concentration changes only slightly during the simulation, since the $c_0$ term is small to have a significant affect in 570 seconds. Thus, for this data set the total receptor dynamics does not affect the LH profile. Receptor desensitization is more prominent with multiple pulses over a longer time period, as we will see in the next section. \\

\begin{table}[b]
\caption{Parameters for Section \ref{midgley}}
\begin{tabular*}
{\textwidth}{@{\extracolsep{\fill} } l l l }
\hline 
$v$=0.2 mm/sec & $k_1$=0.0091 /nM $\cdot$ sec & $a_1$=2.9348 /nM $\cdot$ sec\\
$A$=17 nM & $k_2$=0.0108 /sec & $b_2$=0.0001 /sec \\
$\tau$=240 sec & $a_0$=0.000001 nM/sec & $a_2$=0.0073 /sec\\ 
$R_{in}$=0.2 nM &$b_0$= $a_0$/$R_{in}$ /sec & $b_3$= 0.0402 /sec\\
$bs$=0.0319 nM/sec & $c_0$= 0.0002678 /sec  & $\Delta t$= 0.01 sec\\
$\ell$ =5.56 mm & $b_1$ =0.6058 /sec & $\Delta x$=0.01 mm\\
\hline
\label{table:parametersmidgley} 
\end{tabular*}
\end{table}

\textbf{Determination of Model Parameters}, The details of the experimental setup are explained in \cite{cantor1996use, midgley19958}. The volume of the cell chamber is 32 $\mu$l, with a 2.69 mm diameter and a 5.563 mm height. Thus the area of the cell chamber is 5.6832 mm$^2$. The velocity of the fluid flow is 72 $\mu$l/min. Thus, the one-dimensional velocity $v$ is 0.2 mm/sec. The stimulation time of GnRH, $\tau$, is 240 seconds. The cells are mixed with beads before loading. The beads occupy 17 $\mu$l, whereas the cells and the fluid occupy 15 $\mu$l. Approximately $2 \times 10^5$ cells are loaded into the chamber. We assume there are $10^4$ GnRH receptors per cell \cite{blum2000mathematical}, thus, the initial total GnRH-R concentration is $R_{in}=0.22$ nM. The concentration of the GnRH given is $20$ ng/ml. Assuming the molecular weight of GnRH is $1,183.27$ Daltons, $1$ ng/ml GnRH $=0.84$ nM GnRH. So, the amplitude of the free GnRH is $A=20 $ nM in the simulation. The basal LH level is $26.6$ ng/ml in the data, which is 26.6/30 nM LH, given the molecular weight of LH is 30,000 Daltons. So, $bs=0.0319$ nM/sec. The mathematical model gives LH in nM, however the data is given in ng/ml. Thus, we convert nM LH to ng/ml LH by multiplying the former by 30. The data is the average LH in 30-second fractions, thus we calculate the average LH over 30-second intervals and each average correspond to a black dot in Figure \ref{fig:midgley} Panel A. The simulations run for 570 sec. \\

We determine the other parameters to match the simulation to the experimental data using an educated initial guess and minimizing the squares of errors using built-in MATLAB fminsearch function. The parameters a$_0$ and b$_0$ are chosen to have a constant total receptor concentration in the absence of GnRH. \\

\subsubsection{Trains of Pulses} \label{Mcintosh}

McIntosh et al. \cite{mcintosh1986varying} studied the secretion of LH in response to multiple GnRH pulses with varying pulse durations and periods using a perifusion system and female sheep pituitary cells . They gave multiple pulses and collected the outgoing medium in 6 minute intervals. Note that in Section \ref{midgley} the time is in seconds, whereas in this section it is in minutes. Since the McIntosh data is not as finely sampled as the previous data, they did not observe the triphasic response.\\

\begin{figure}[t] \centering 
\includegraphics[width=\textwidth]{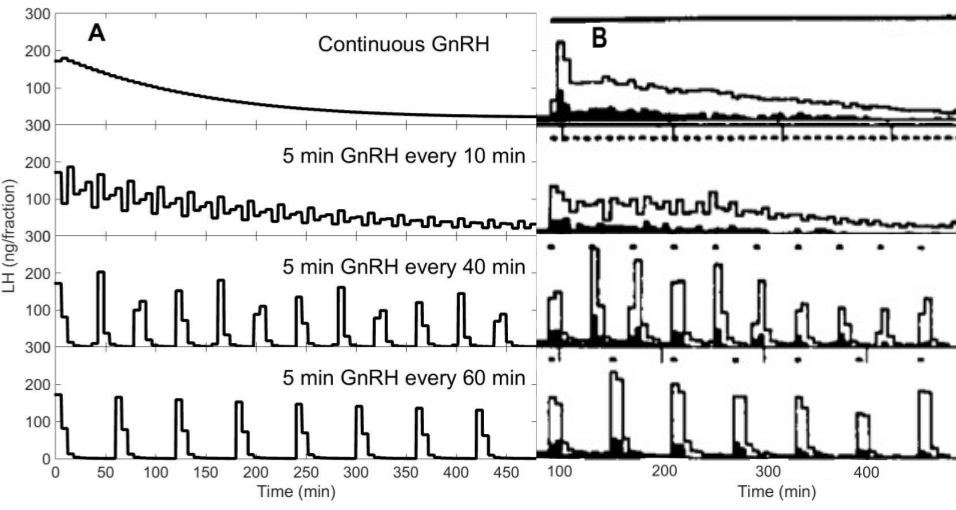} 
\caption{ {\footnotesize{\textbf{LH secretion patterns.} Panel A shows the simulation results; Panel B shows the corresponding data. The black lines at the top of the data show GnRH stimulation pattern, which are written at the left top corner of the simulation results. First row demonstrates continuous GnRH stimulation, the second row 5 minute GnRH pulses every 10 minutes, the third row 5 minute GnRH every 40 minutes and the last row shows 5 minute GnRH pulses every 60 minutes. The simulations run for 480 minutes and the medium coming from the cell chamber is collected for 6 minutes. For the graphs on Panel B, the black lines representing GnRH stimulation patterns are at the level of 300 ng/fraction on the y-axis. The data is modified from \cite{mcintosh1986varying}.}}}
\label{fig:mcintosh}
\end{figure}

\begin{figure} \centering 
\includegraphics[width=\textwidth]{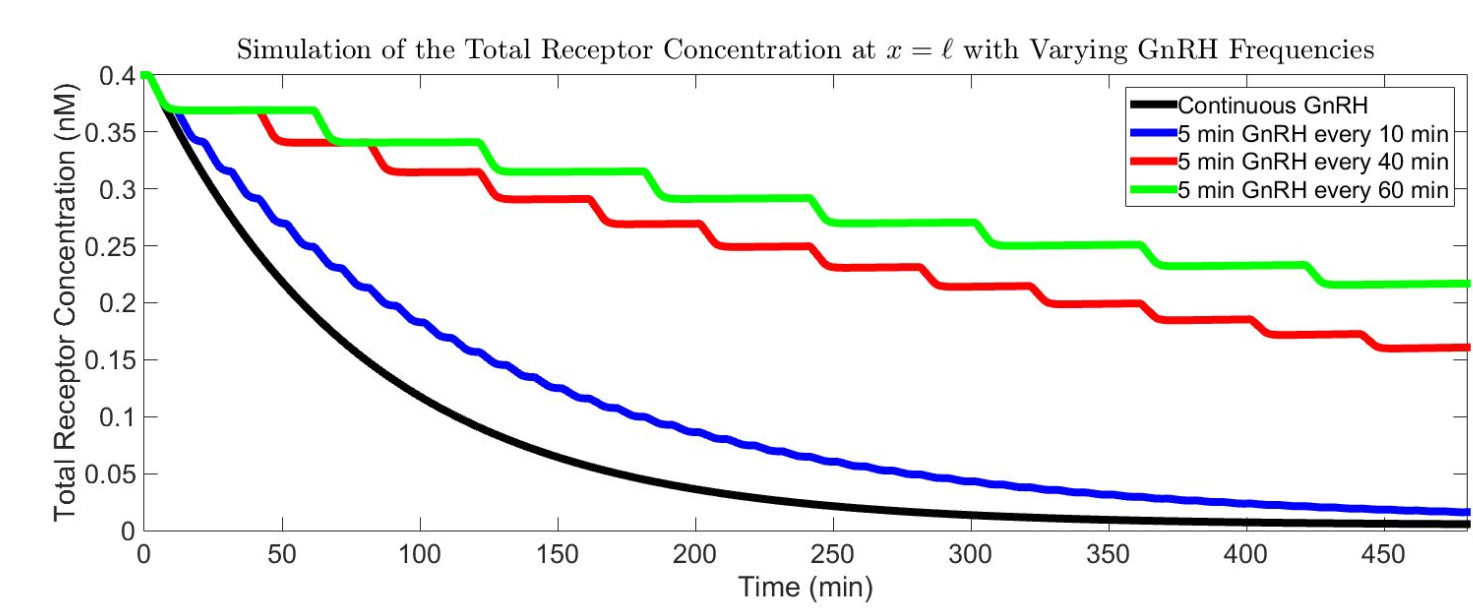} 
\caption{{\footnotesize{\textbf{Simulations of the Time Courses of Total Receptor Concentrations at $x=\ell$.} The total receptor concentration decreases with every GnRH pulse. Four different cases are given: continuous GnRH stimulation (black); 5 min every 10 min (blue); 5 min every 40 min (red); 5 min every 60 min (green). The simulation is run for 480 minutes.}}}
\label{fig:total receptor}
\end{figure}

\textbf{Model Simulation}, Figure \ref{fig:mcintosh} shows the response of the gonadotroph cells to a variety of GnRH stimulation patterns. Panel A shows the simulation results and Panel B shows the data. The amount of LH in each fraction is marked as a line over the time interval the fraction is collected. In the data open rectangles are LH, whereas filled rectangles are FSH. We are only interested in the LH secretion in this work. The black bars at the top of the data show GnRH stimulation patterns. In Panel B, as in Panel A, each of the y-axes goes from 0 to 300. Notice that in all these cases the simulation outcomes are very similar to the experimental results.\\

In Figure \ref{fig:mcintosh}, the first row shows continuous GnRH stimulation. In the following rows, GnRH is given for 5 minutes every 10 minutes, every 40 minutes and every 60 minutes. Continuous stimulation (first row) and five-minute pulses every 10 minutes (second row) eventually suppress the secretion of LH both in the simulation and in the data. In the other two cases, the peaks decrease in height gradually with consequent pulses. This diminishing of LH response happens due to the receptor desensitization, which is represented by the $c_0B(x,t)$ term in the equation \eqref{eq:R}. In Figure \ref{fig:total receptor} the total receptor concentrations at $x=\ell$ for the four different cases are presented. With every pulse, the bound GnRH leads to a decrease in the total receptor concentration. In continuous and high frequency stimulations, the total receptor concentrations decrease rapidly as seen by the black and blue curves. However, in the other two cases, fewer pulses are given, and thus the total receptor concentration decreases more slowly as seen in the red and green curves.\\

\textbf{Importance of The Length of the Collection Time Interval}, The simulations of LH at $x=\ell$ generate a continuous and periodic LH pattern with the same shaped response for every pulse, yet, with decreasing heights due to the receptor desensitization. However, integrating this regular LH output over a fraction collection time which does not divide the period of the simulations evenly leads to an irregularly shaped LH outcome with varying heights as in Figure \ref{fig:mcintosh}, third row of Panel A, where 5 minute GnRH pulses are introduced every 40 minutes and the medium is collected in 6 minute fractions. The LH output for this case has three different shapes (compare the second, third and fourth pulses) with different heights repeated for the rest of the simulation. This pattern looks like an interesting biological phenomenon, however, this irregularity in the simulation is only an artifact of the fraction collection time. Unlike 40-minute stimulation period, the stimulation with period 60 minutes has the same shape repeated (fourth row), since 6 minute fraction collection time divides 60 minute period evenly. Thus, the length of the collection time interval determines the regularity and the height of the LH output. \\ 

The difference in the first two LH peaks of the data of 40 minute pulse period (Panel B, third row) is partly due to the way the medium is collected, but also self-priming might play a role (See Discussion). \\

\begin{table}[b]
\caption{Parameters for Section \ref{Mcintosh}}
\begin{tabular*}
{\textwidth}{@{\extracolsep{\fill} } l l l }
\hline 
$v$= 7 mm/min  & $k_1$ = 60 $\times$ 0.0091 /nM $\cdot$ min & $a_1$ = 60 $\times$ 2.9348 /nM $\cdot$ min\\
$A$= 4.23 nM & $k_2$ = 60 $\times$ 0.0108 /min & $b_2$ = 60 $\times$ 0.0001 /min\\
$\tau$ = varying &$a_0$ = 60 $\times$ 0.000001 nM/min & $a_2$ = 60 $\times$ 0.0073 /min \\ 
$R_{in}$ = 0.4 nM & $b_0$ = $a_0$/$R_{in}$ /min & $b_3$ =  60 $\times$ 0.0402 /min\\
$bs$ = 0 nM/min & $c_0$ = 60 $\times$ 0.0002678 /min  & $\Delta t$ = 0.001 min \\
$\ell$ = 10 mm & $b_1$ = 60 $\times$ 0.6058 /min & $\Delta x$ = 0.01 mm\\
\hline
\end{tabular*}
\label{table:parametersmcintosh}
\end{table}

\textbf{Determination of Model Parameters}, The flow rate is 0.14 ml/min. Approximately $5 \times 10^6$ cells are loaded to the columns. GnRH concentration is 4.23 nM. They use 1 ml syringe barrels as the cell chamber \cite{mcintosh1984microcomputer, murray1994modeling}. Although they don't give the area of the particular syringe they used, we assume the area is about 20 mm$^2$ based on the average 1 ml syringe dimensions. Thus, the one-dimensional velocity is 7 mm/min. In the diagrammatic description of the perifusion apparatus in \cite{murray1994modeling} the volume of the column contents is given as 0.4 ml, where 0.2 ml of it is occupied by the packing material, leaving the cells 0.2 ml. So, the height of the volume that the cells cover is 10 mm. We take $bs=0$, because the basal level is indistinguishable in this data set. They collect 6-minute fractions, and give their results as ng/fraction. We change nM to ng/ml by multiplying LH by 30 and change ng/ml to ng/fraction by further multiplying the result by 0.84. Finally, we take the average of LH over 6 minute time periods. Unlike the previous part, the averages are not given as dots, but presented as bars over the time period the average is taken. \\ 

This data sets serves as a cross validation of the parameters used in the Section \ref{midgley}. We use the same physiological parameters multiplied by 60 to convert /sec to /min. \\

\subsubsection{Oxytocin vs GnRH} \label{Tabak}

Gonzales-Iglesias et al. \cite{gonzalez2014direct} compared the effects of oxytocin and GnRH on adult female Sprague Dawley rat pituitary cells. Both hormones lead to secretion of LH, though with different dynamics.  \\
  
\begin{figure} [t] \centering 
\includegraphics[width=\textwidth]{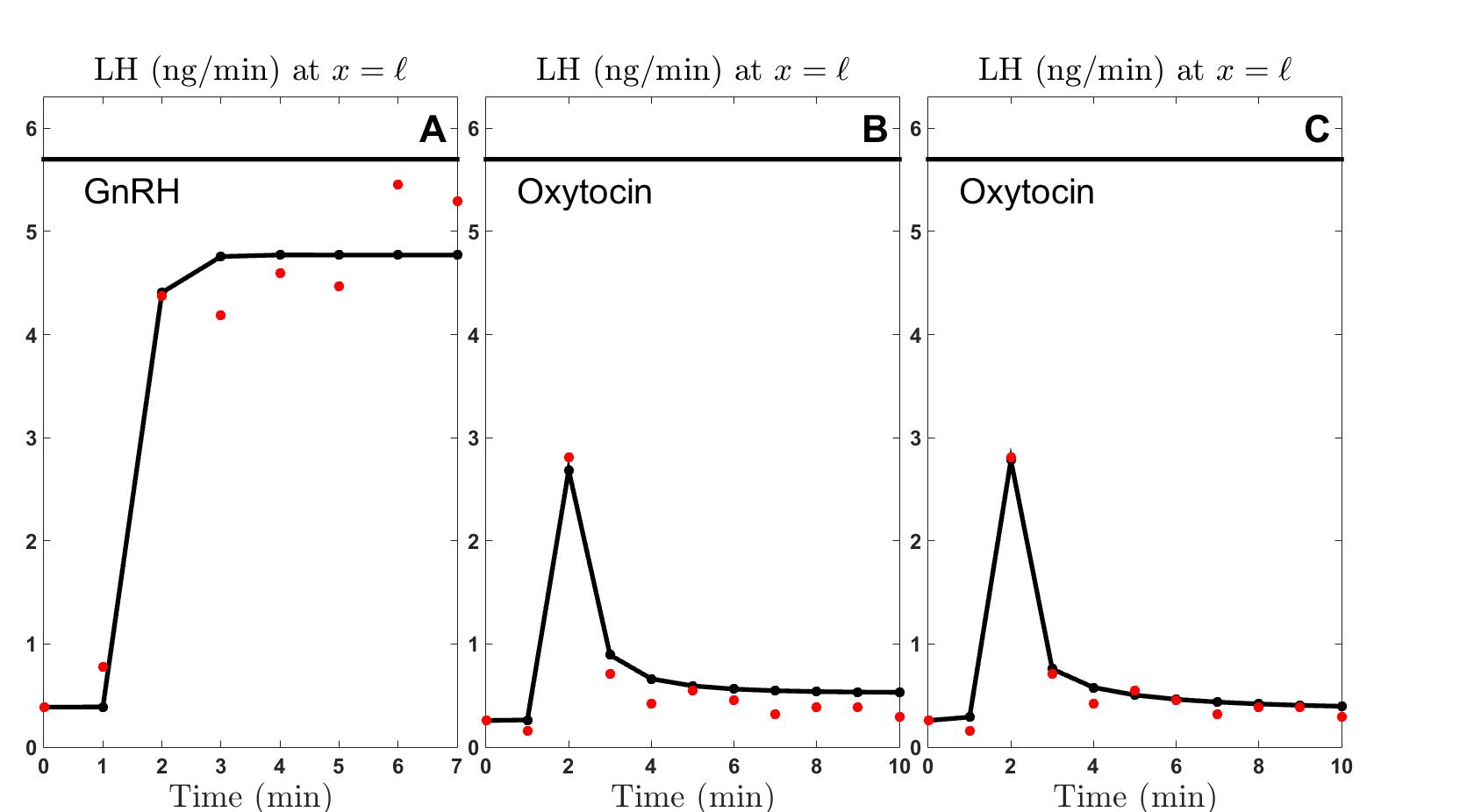} 
\caption{{\footnotesize{\textbf{LH Response to GnRH and Oxytocin.} The red points are the data taken from \cite{gonzalez2014direct}; the black dots and lines are the simulation results. Panel A shows LH secretion in response to 7 mimutes 5 nM GnRH stimulation. Panels B and C show the LH response to 10 nM oxytocin; simulation time is 10 minutes. In the middle graph only the parameter $c_0$ is different, whereas in the right graph both $c_0$ and initial total receptor concentration are different than the parameters used in the GnRH stimulation graph.}}}
\label{fig:tabak}
\end{figure}

\textbf{Simulation Results}, Figure \ref{fig:tabak} shows the LH response to 7 minutes 5 nM GnRH stimulation in Panel A and to 10 minute 10 nM oxytocin stimulation in Panels B and C. The red dots are the data points taken from \cite{gonzalez2014direct}, whereas the black dots and the black lines connecting them are simulation results. Oxytocin stimulation leads to a lower level of LH secretion compared to GnRH stimulation. \\ 

In all the graphs there is no difference in the LH content of the first two fractions, since the passage time through the chamber is one minute, given that the velocity is 3.8 mm/min and the chamber length is 3.8 mm.\\ 

We match the parameters to reproduce the GnRH data and then recover oxytocin data by changing only the parameters related to the receptor dynamics. The strength of the receptor desensitization is controlled by the parameter $c_0$ as represented in the equation \eqref{eq:R}. Since the oxytocin receptor desensitizes faster than the GnRH-R, $c_0$ term for the oxytocin receptor should be larger (see Discussion). In Figure \ref{fig:tabak} Panel B , all the parameters, except $c_0$, are the same as the ones used for Panel A. Also, the densities of the GnRH and oxytocin receptors might be different on the cell membrane. In Panel C, all the parameters, except $c_0$ and $R_{in}$, are the same as the ones used for Panel A. By changing two parameters instead of one, the fit of the simulation to data is improved. The receptor desensitization and/or the difference in the receptor concentration on the cell membrane can explain the difference in the action of these two hormones.\\

\textbf{Determination of Model Parameters}, Approximately $4 \times 10^6$ cells were put into 0.5 ml chambers with 13 mm diameter (personal communication). Thus $\ell$ is approximately 3.8 mm.  They give 5 nM GnRH for 7 minutes and 10 nM oxytocin for 10 minutes. They collect the medium at 1-minute intervals. Flow rate is 0.5 ml/min, thus one-dimensional velocity is 3.8 mm/min. Assuming that there are about $10^4$ GnRH receptors per cell, the concentration of GnRH receptors in the cell chamber is approximately 0.13 nM. The results are given in ng/min. In 1 min they collect 0.5 ml, so we multiply LH first by 30 to convert nM to ng/ml, then by 0.5 to convert ng/ ml to ng/min. The parameter $bs$ is calculated using the first point in the data sets, which are 0.3871 ng/min for GnRH stimulation data and 0.2581 ng/min for oxytocin stimulation data, which correspond to $bs=0.0258$ nM/min and $bs=0.0172$ nM/min respectively. The rest of the parameters are determined to match the data as in the previous sections.\\

To reproduce the difference in GnRH stimulated LH release and oxytocin stimulated LH release, first we only changed $c_0$ parameter from 0.0000916  /min to 232 /min in Figure \ref{fig:tabak} Panel B. In Panel C we change both $c_0$ and $R_{in}$: we change $c_0$ from 0.0000916  /min to 1550 and $R_{in}$ from 0.13 nM to 0.95 nM.\\

\begin{table}[b]
\caption{Parameters for GnRH Stimulation Section \ref{Tabak}}
\begin{tabular*}
{\textwidth}{@{\extracolsep{\fill} } l l l }
\hline 
$v$ = 3.8 mm/min & $k_1$ = 0.072 /nM $\cdot$ min & $a_1$ =1866 /nM $\cdot$ min    \\
$A$ = 5 nM & $k_2$= 27.213 /min & $b_2$ = 0.0187 /min\\
$\tau$ = 7 min & $a_0$ =0.001494 nM/min & $a_2$= 4.3148/min\\
$R_{in}$ = 0.13 nM & $b_0$ = $a_0$/$R_{in}$ /min &  $b_3$ = 0.5449 /min \\ 
$bs$ = 0.0258 nM/min &  $c_0$= 0.0000916  /min & $\Delta t$ = 0.001 min\\ 
$\ell$ = 3.8 mm  &  $b_1$ = 1368  /min & $\Delta x$ = 0.01 mm\\
\hline
\end{tabular*}
\label{table:parameterstabak} 
\end{table}

\subsection{In silico experiments} \label{insilico}

\subsubsection{Velocity Dependence of Data} \label{vdep}

\begin{figure}[t] 
\centering
\includegraphics[width=\textwidth]{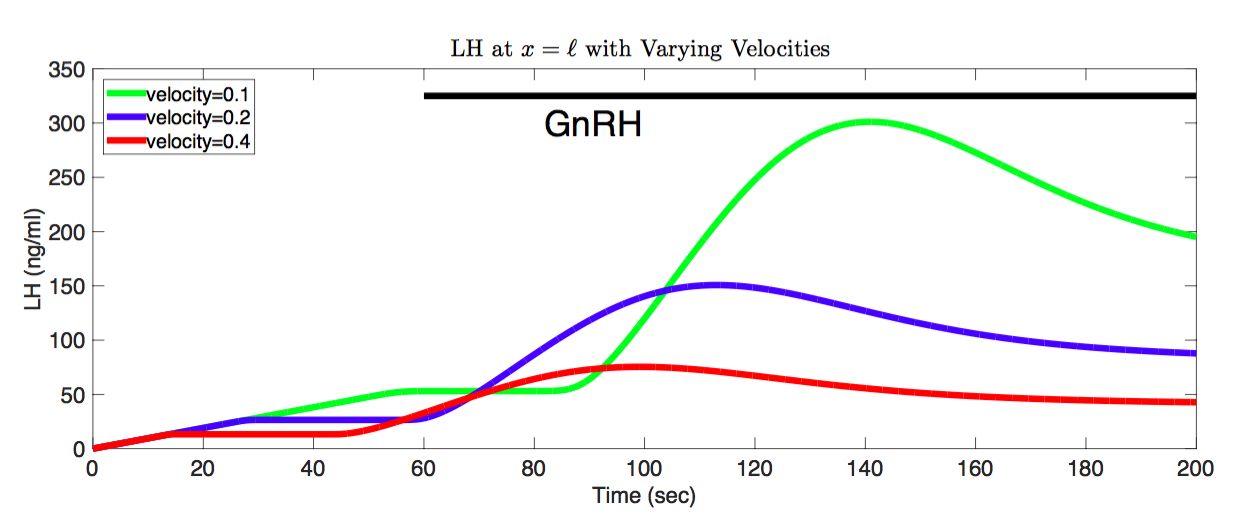} 
\caption{{\footnotesize{\textbf{Dependence of LH at $x=\ell$ on medium flow velocity.} The green line shows the simulation with medium flow velocity 0.1 mm/sec, the blue line with 0.2 mm/sec and the green line with velocity 0.1 mm/sec. 17 nM GnRH is supplied to the cell chamber starting at the 60th second for the rest of the simulation, represented by the black line at the top. }}} \label{fig:vdep}
\end{figure} 

To examine the influence of the medium flow velocity on the outcome of the perifusion experiments, we use the parameters from Section \ref{midgley} and vary the velocity. Figure \ref{fig:vdep} shows the LH at $x=\ell$ with three different velocities. The red line shows the simulation with velocity 0.4 mm/sec, the blue line with velocity 0.2 mm/sec and the green line with velocity 0.1 mm/sec. \\

As explained in Section \ref{mathematicalmodel}, the steady state basal LH level at $x=\ell$ is equal to $(bs \times \ell)/v$. Thus, small velocities lead to a higher steady state basal LH level, as shown in the Figure \ref{fig:vdep}. The smallest velocity (green) has the highest steady state LH level. More specifically, since $\ell=5.56$ mm, $v=0.1$ mm/sec, $bs=0.0319$ nM/sec, the steady state basal LH level at $x=\ell$ is 53.2 ng/ml. This steady state LH level is 26.6 ng/ml for the velocity 0.2 mm/sec (blue), and 13.3 ng/ml for the velocity 0.4 mm/sec (red). Also, slow velocities take longer to reach this steady state basal LH level. The time required to reach this basal level is $ \ell/v$ , which is 55.6 seconds for velocity 0.1 mm/sec; 27.8 mm/sec for the velocity 0.2 mm/sec and 13.9 mm/sec for the velocity 0.4 mm/sec. \\ 

In the simulations, the GnRH is supplied into the cell chamber starting at 60th second until the end of the simulation, as represented by the black line at the top of Figure \ref{fig:vdep}. The LH levels at the right end point of the chamber start increasing after $\ell/v$ seconds passage time through the chamber. This passage time is 55.6 seconds when velocity is 0.1 mm.sec, 27.8 seconds when velocity is 0.2 mm/sec and 13.9 mm/sec when the velocity is 0.4 mm/sec. When velocity is small, the quantity of the LH collected at the right end point of the chamber is greater, since with smaller velocities more LH accumulates in the chamber, before it is washed out. Thus, when the velocity is small, the LH appears later in the collections, however, in greater quantities.\\

\subsubsection{Movement of GnRH in the Cell Chamber} \label{GnRHmovement}

\begin{figure}[t] 
\centering
\includegraphics[width=\textwidth]{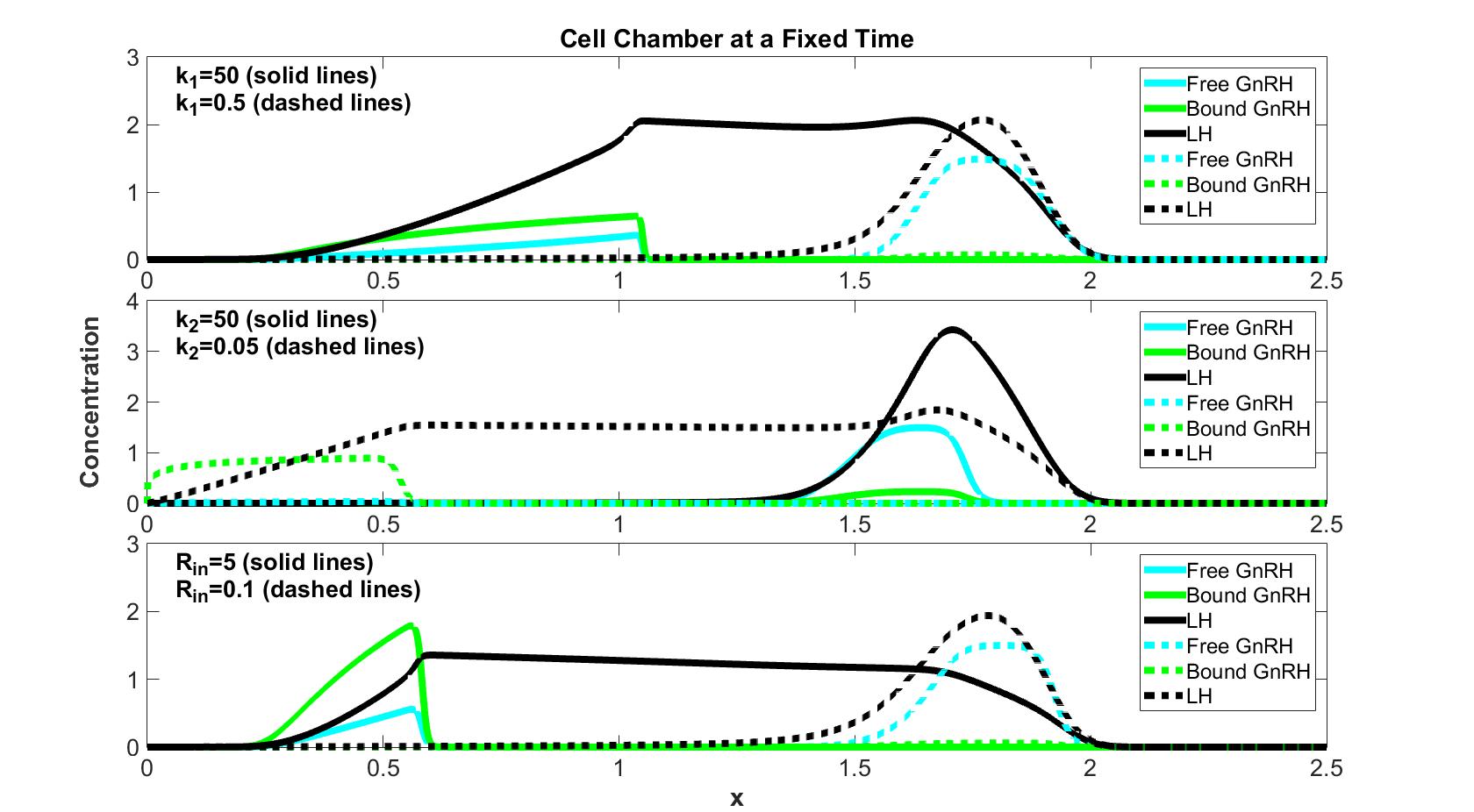} 
\caption{{\footnotesize{\textbf{GnRH Movement in the Cell Chamber with Varying Parameters.} The simulations show free GnRH (cyan lines), bound GnRH (green lines) and LH (black lines) concentrations at a fixed time in the cell chamber. In each row the solid lines and the dashed lines give the results of two simulations where all the parameters are the same except one key parameter.}}} \label{fig:gnrhmovement}
\end{figure}

In this section we give a qualitative description of how the GnRH movement down the chamber depends on the binding rate of free GnRH to the receptor, $k_1$; the dissociation rate of bound GnRH from the receptor, $k_2$ and the initial GnRH-R concentration, $R_{in}$. Each row in Figure \ref{fig:gnrhmovement} shows two simulations in which only one parameter is changed. In the top row, all parameters except $k_1$ are the same; in the middle row all parameters except $k_2$ are the same and in the bottom row all parameters except the initial receptor concentration $R_{in}$ are the same. The black lines show LH, the green lines show bound GnRH and the cyan lines show free GnRH in the chamber at $t=20$ seconds. The common parameters used in all simulations shown in Figure \ref{fig:gnrhmovement} are presented in the Table \ref{table:movementGnRH}. \\ 

The fronts of both the solid and dashed black lines are at the same point in all simulations, since the movement of the LH down the chamber does not depend on $k_1$, $k_2$ or $R_{in}$. Once LH is secreted from the cells into the chamber, it moves with the velocity of the fluid flow, $v$. Also, in all simulations the free and bound GnRH move together. \\ 

The first row in Figure \ref{fig:gnrhmovement} shows the impact of $k_1$ on the GnRH movement. For this row $k2=10$ /min and $R_{in}=1$ nM. The solid lines show the simulation with $k_1=50$ /nM $\cdot$ min and the dashed lines show the simulation with $k_1=0.5$ /nM $\cdot$ min. When $k_1$ is large (solid lines), free and bound GnRH move more slowly and the bound GnRH level (green) is high. Thus, large $k_1$ leads to stronger activation of the signaling cascade, and eventually it leads to higher LH secretion. When comparing the total LH secreted, we compare the areas under the LH curves. \\ 

The middle row in Figure \ref{fig:gnrhmovement} illustrates the influence of $k_2$ on GnRH movement. For these simulations $k_1=10$ /nM $\cdot$ min; $k_2=50$ or $=0.05$ /min and $R_{in}=1$ nM. When $k_2$ is large (solid lines), free and bound GnRH move faster. The rate of dissociation affects the tail of the GnRH bumps: if $k_2$ is high, the GnRH bumps are narrow (solid lines) and when $k_2$ is low, the bumps have longer tails (dashed lines). Thus, with small $k_2$ the bound GnRH dissociates more slowly, so the signaling cascade is activated for a longer time period. In our simulations with the particular choice of parameters, $k_2$ determines the rate at which LH at $x=\ell$ decreases to its basal level after the termination of free GnRH introduction into the chamber. \\ 

The bottom row in Figure \ref{fig:gnrhmovement} shows two simulations with different $R_{in}$ values, where $k_1=10$ /nM $\cdot$ min, $k_2=10$ /min and $R_{in}=5$ or 0.1 nM. In this section the parameters $a_0$, $b_0$ and $c_0$ are zero, thus receptor concentrations are constant throughout the simulations. When receptor concentrations are high (solid lines), bound GnRH concentrations are high (green) and GnRH moves more slowly.\\ 

To sum up, when $k_1$ is high, $k_2$ is small and/or the receptor concentrations are high, GnRH moves more slowly, bound GnRH concentrations are higher, and more LH is produced in the chamber. \\

\begin{table}[]
\caption{Parameters for Section \ref{GnRHmovement}}
\begin{tabular*}
{\textwidth}{@{\extracolsep{\fill} } l l l }
\hline 
$v$ = 0.1 mm/min & $k_1$ = 10 /nM $\cdot$ min & $a_1$ =10 /nM $\cdot$ min \\
$A$ = 1.5 nM & $k_2$=10 /min & $b_2$ =0.4 /min\\
$\tau$ = 3 min & $a_0$=0 nM/min& $a_2$=0.5 /min\\
$R_{in}$ = 1 nM & $b_0$ = $a_0$/$R_{in}$ /min &$b_3$ =0.5 /min \\ 
$bs$ = 0 nM/min & $c_0$= 0  /min & $\Delta t$ = 0.001 min \\
$\ell$ = 2.5 mm & $b_1$ = 3 /min & $\Delta x$ = 0.001 mm\\
fixed time = 20 min &  &\\
\hline
\end{tabular*}
\label{table:movementGnRH} 
\end{table}

\subsubsection{Dependence of LH Secretion Efficiency on Pulse Characteristics}\label{specificresponse}

\begin{figure}[t] \centering 
\includegraphics[width=\textwidth]{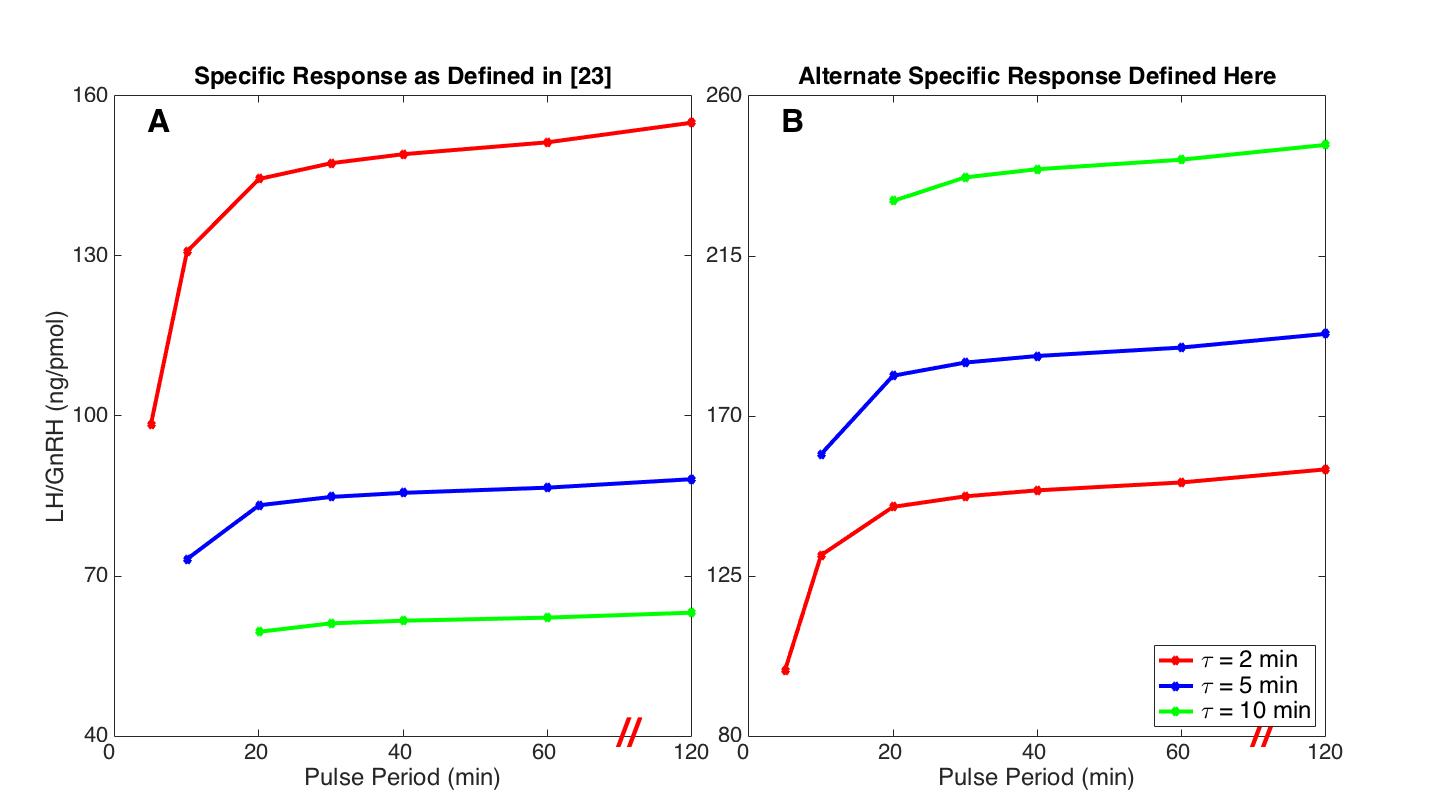} 
\caption{{\footnotesize{\textbf{Specific Response with Varying Pulse Characteristics.} The x-axes show pulse period in minutes; the y-axes show the specific response in ng/pmol. Three different pulse durations, $\tau$, are considered: 2 minutes (red), 5 minutes (blue), and 10 minutes (green). The average LH output per pulse using the second, the third and the fourth pulses is calculated and then divided by the total GnRH amount supplied in one pulse. Panel A shows the case where 4.23 nM GnRH is given with each pulse for every pulse duration resulting in 4.23 $\times \tau$ nM$\cdot$min total GnRH per pulse. In Panel B, 4.23 nM GnRH is given for $\tau=2$ minutes, 1.692 nM GnRH for $\tau=5$ minutes, and 0.846 nM GnRH for $\tau=10$ minutes, so that the total amount of GnRH given in each pulse is the same, 8.46 nM$\cdot$min. Note that in Panels A and B the scales of y-axes are different and the red lines are the same. Long pulse durations are less efficient if the same GnRH concentration is supplied (Panel A); however, they are more efficient if the total GnRH amount per pulse is kept constant (Panel B).}}}
\label{fig:specific response}
\end{figure}

To analyze further the effects of the pulse duration and period, McIntosh et al. \cite{mcintosh1986varying} calculated the LH output per unit of GnRH introduced into the chamber for different pulse characteristics, which they called the ``specific response''. To this end, they calculated the average LH output per pulse using the second, the third and the fourth pulses and then divided it by the total amount of GnRH given in each pulse. In their experiments the total amount of GnRH per pulse depends on the pulse duration $\tau$. We calculated their specific response in our simulations; the results are presented in Figure \ref{fig:specific response} Panel A. The x-axis shows the pulse period in minutes and the y-axis shows the specific response in ng/pmol. The red line shows the specific response when the pulse duration $\tau=2$ minutes, the blue line when $\tau=5$ minutes and the green line when $\tau=10$ minutes. The parameters used in the simulations are from Section \ref{Mcintosh}. For the pulse length $\tau$ specified, 4.23 nM GnRH is supplied to the chamber resulting in 4.23$\times \tau$ nM$\cdot$min total GnRH per pulse. For a fixed pulse period, the specific response is higher when the pulse duration is shorter. For a fixed pulse duration $\tau$, as the time interval between the pulses increases, the specific response increases. These results reproduce qualitatively Figure 4 in \cite{mcintosh1986varying}. \\ 

Alternatively, one could calculate a different ``specific response'' by giving the same total GnRH in every pulse independent of the pulse duration $\tau$. To calculate the specific response with this alternative way, we simulated the experiment with 4.23 nM GnRH for the pulse duration 2 min, or 1.692 nM GnRH for 5 min or 0.846 nM for 10 min, so that each pulse delivers 8.46 nM $\cdot$ min GnRH. Then we calculated the specific response by calculating the average LH output per pulse using the second, the third and the fourth pulses and then dividing it by the 8.46 nM $\cdot$ min total GnRH amount. The results are presented in Figure \ref{fig:specific response}, Panel B. Unlike Panel A, with this alternative definition, the specific response was higher with longer pulse duration $\tau$. Observe that in Figure \ref{fig:specific response} the Panels A and B have different y-axis scales and the red lines corresponding to $\tau=2$ minutes are the same lines.\\

Since the LH amount a cell can secrete is limited, using high GnHR concentrations does not lead to a proportional increase in the LH amount secreted. When 4.23 nM or 0.846 nM of GnRH is given for 10 minutes, the LH output at $x=\ell$ shows a triphasic response, where it first peaks and then decreases down to a quasi steady state level until the end of stimulation. The quasi steady state LH level at $x=\ell$ is higher with 4.23 nM GnRH compared to 0.846 nM GnRH, but not 5 times higher. Thus, dividing the average LH output per pulse by the total GnRH given to the chamber resulted in a lower specific response with the 4.23 nM stimulation compared to 0.846 nM. \\

Although McIntosh et al concluded that with decreasing pulse duration the responsiveness of the cells is increased, our simulations suggest that long pulse durations with low GnRH concentrations lead to higher LH output per GnRH amount supplied.\\

\section{SENSITIVITY ANALYSIS}

\begin{figure}[t] 
\centering
\includegraphics[width=\textwidth]{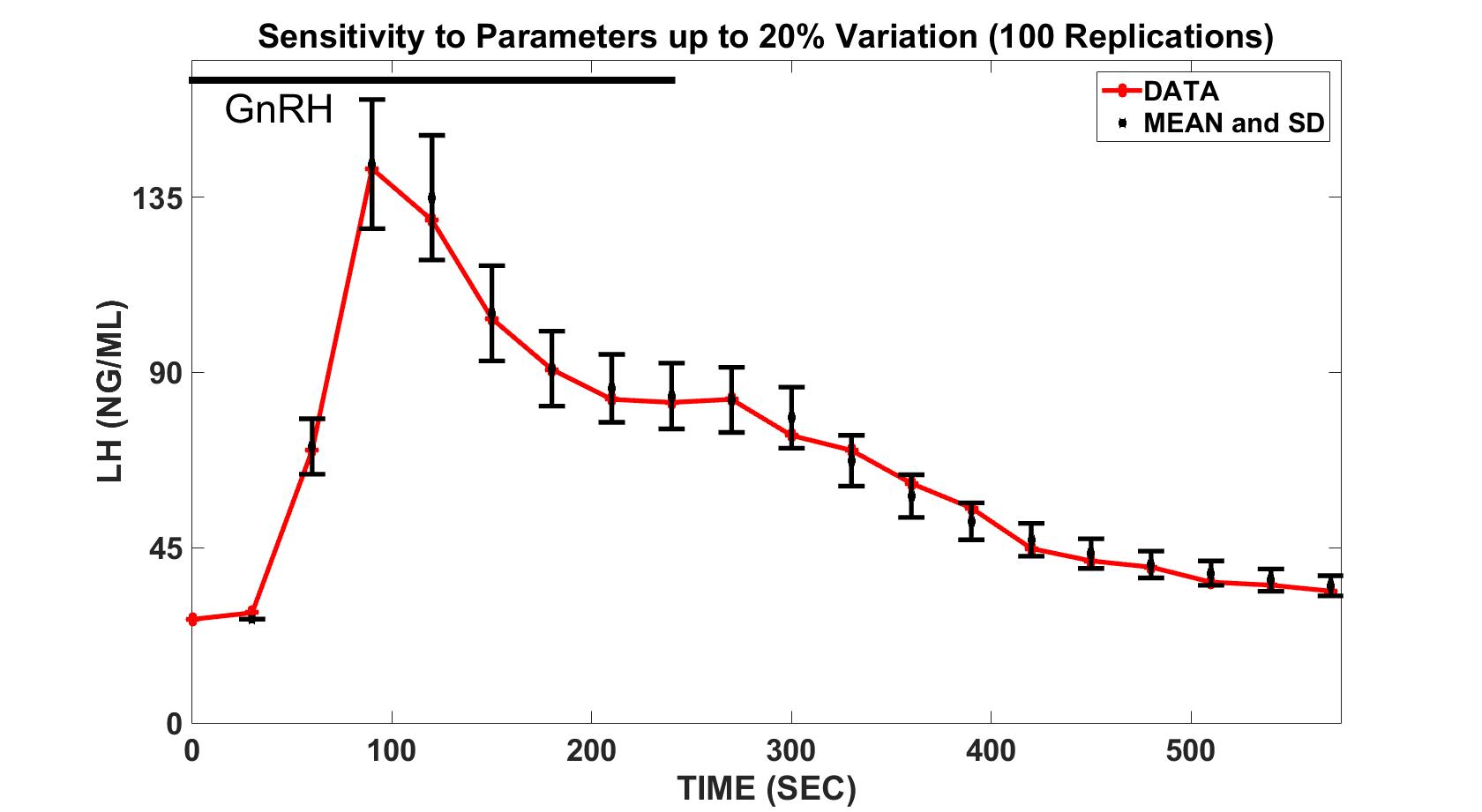} 
\caption{\footnotesize{\textbf{Sensitivity of LH outcome at $x=\ell$ to parameters.} The parameters used for Section \ref{midgley} are varied randomly up to 20$\%$ and 100 simulations are run with perturbed parameters. The mean and the standard deviation for these 100 runs are calculated and plotted with the data. The red dots and the line connecting them is the data from \cite{cantor1996use} and the black points and the black bars are the mean and the standard deviation of the 100 simulations with perturbed parameters. The black bar at the top of the graph shows the GnRH stimulation. We see that the characteristic triphasic response is not sensitive to parameter choices and is preserved over a range of parameters.}} 
\label{fig:sensitivity}
\end{figure}

To investigate how sensitive the characteristic triphasic response is to choices of parameter values, we vary the parameters up to 20 $\%$. For every parameter we create a region with 20  $\%$ lower and 20  $\%$ higher than the particular parameter value and then we randomly pick the parameter from that region with a uniform distribution. We perturb all the parameters at the same time and generate 100 runs with varied parameter and for each data point we calculate the mean and the standard deviation. In Figure \ref{fig:sensitivity} the red line shows the data from \cite{cantor1996use} and the black bars are centered at the mean of the 100 runs, and the bars show the region within one standard deviation of the mean. The black bar at the top of the graph shows GnRH stimulation. The triphasic response is preserved even with the perturbed parameters, indicating that this characteristic shape of the response is not sensitive to parameters chosen. \\

\section{DISCUSSION}
Many hormones are secreted in pulses including GnRH, growth hormone, adrenocorticotropic hormone, oxytocin, insulin and glucagon \cite{veldhuis2008motivations}. Perifusion cell cultures permit the study of such intermittent stimulation in a controlled environment. Mathematical models are crucial tools for interpreting the results of these perifusion experiments and for connecting their outcomes to cellular events. \\ 

Systems of ordinary differential equations model well static cell cultures, where cells are incubated in a well-mixed stationary medium. However, in perifusion cell cultures the medium is not homogeneous throughout the chamber and is constantly moving. The secreted products and the signal introduced into the chamber are constantly washed out. Thus, in order to interpret the perifusion data one must consider the spatial aspect of these experimental systems. In this paper we developed and analyzed a mathematical model of coupled partial differential equations for perifusion cell culture experiments, which combined the movement of the substances in the chamber with the intracellular events leading to LH secretion.\\ 

In Section \ref{perifusion} we matched the model to three different data sets and the model fit well to all of them. In Section \ref{midgley} we determined the parameters to reproduce the data from \cite{cantor1996use} and in Section \ref{Mcintosh} we used the same parameters to reproduce the experiments from \cite{mcintosh1986varying}. Thus, the second set of data served as a cross validation for the choice of parameters. For the third data set from \cite{gonzalez2014direct} we used a different set of parameters. That is biologically reasonable since the experimenters used a different cell type than the other two groups. \\   

In Section \ref{midgley} we analyzed the triphasic LH response at $x=\ell$, which consisted of the LH peak, followed by the lower quasi steady state level and the decrease back to the basal secretion level. The negative feedback from the phosphatases was crucial in reaching the lower quasi steady state LH level at $x=\ell$. The desensitization of the receptors did not have a significant effect, since the data was collected over a relatively short time interval. In our model, the rate of return to the basal LH level at $x=\ell$ depended primarily on the dissociation rate of bound GnRH from its receptor, $k_2$. \\ 

In Section \ref{Mcintosh} the data from \cite{mcintosh1986varying} was collected over 480 minutes. The receptor desensitization led to the eventual suppression of the LH secretion in this prolonged GnRH exposure. The first two data sets show that in our model the phosphatase based negative feedback is fast and important over short time intervals, whereas GnRH receptor desensitization affects the outcome over long time intervals \cite{mcardle2002signalling}. Note that in the third row of Figure \ref{fig:mcintosh}, the second GnRH pulse leads to a higher response than the first pulse both in the simulation and in the data. Also the heights of the LH outcome vary throughtout the GnRH stimulation. In the simulation, this variation depends solely on the fraction collection time: fraction collection times that do not divide the period of the stimulation evenly lead to irregularities in the data. To differentiate between the artifacts of sampling and interesting biological phenomena, the medium collection time should be chosen so that it divides the period of the stimulation evenly and it should be fine enough to capture the different phases of the LH secretion pattern. \\

Some irregularity in the data might also come from priming. Increased sensitivity of gonadotrophs to subsequent GnRH pulses is called self-priming \cite{leng2008priming}. Our model lacks the biological mechanisms which might be responsible for priming such as the integration of the fast Ca-mediated pathway and the slow cAMP-mediated processes \cite{evans2013two,evans1999modulation}, exposure to estrogen \cite{leng2008priming}, the autocrine affects of pituitary derived GnRH \cite{krsmanovic2000local}, and repositioning of secretory granules to cell membrane through microfilament reorientation \cite{scullion2004modelling}. \\

The third data set from \cite{gonzalez2014direct} compared the GnRH and oxytocin stimulation in rat pituitary cells. GnRH and oxytocin receptors are both G-protein coupled receptors connected to the same downstream effectors, however, unlike GnRH receptors, oxytocin receptors can undergo rapid desensitization \cite{gimpl2001oxytocin}. After choosing the parameters to match our model to GnRH stimulation data, we reproduced the oxytocin stimulation results by altering only one or two parameters related to receptor dynamics and keeping the rest of the parameters the same. Thus, we showed that different receptor dynamics might underlie the difference in the GnRH and oxytocin experimental results.\\

In Section \ref{insilico}, we conducted \textit{in silico} experiments to explore the importance of some key parameters. In Section \ref{vdep}, we showed that with slow velocities the time required to reach the steady state basal LH level at $x=\ell$ is longer and the LH content of the outgoing medium is higher. One might think that in case of slow velocities the autocrine signals are not washed out, leading to higher secretion \cite{krsmanovic2000local}. To differentiate between biological effects and the influence of medium flow velocity, one should calculate the increase in the LH levels due to lowering the medium flow velocity and if the effect seen is more prominent than expected, then one should look for biological explanations. \\ 

In Section \ref{GnRHmovement} we focused on the movement of GnRH down the chamber. High binding rate, $k_1$, low dissociation rate, $k_2$, and high total GnRH receptor concentration lead to high bound GnRH levels, thus more production of LH, and also slower movement of GnRH down the column. This suggests an inverse mathematical problem and a possible technique for determining binding and dissociation constants: the substance whose binding and dissociation rates are to be determined can be passed through a column with a known receptor concentration. By carefully sampling the outflow, one can determine the rate constants. This approach was used in \cite{reed1994mathematical} and \cite{reed1990approximate}.\\ 

In Section \ref{specificresponse} we investigated the dependence of the LH secretion amount on the pulse characteristics. In  \cite{mcintosh1986varying} the notion of ``specific response'' was introduced, which is calculated by dividing the average LH output per pulse by the total amount of GnRH introduced in one pulse. They found that short pulse durations $\tau$ with long in between pulse intervals was the most efficient way of stimulation. Our simulations reproduced their results. However, we suggested an alternative definition of ``specific response''. In \cite{mcintosh1986varying} the total amount of GnRH introduced into the chamber in one pulse depended on the pulse duration $\tau$, where long pulses resulted in high total GnRH amount per pulse. In our alternative definition we keep the GnRH amount per pulse independent of $\tau$. Our simulations indicate that longer pulses with lower GnRH concentrations are more efficient than high concentration short duration pulses.\\ 

Our mathematical model can be expanded to incorporate more downstream elements relevant to the production of LH, for example the IP$_3$ and DAG pathway \cite{ruf2004genomics}. Also, our methods can be used to understand perifusion experiments for other hormones by including relevant cellular mechanisms for their signaling pathways. In addition, for experimental systems with very slow velocities or small $\ell$, a Laplacian term can be added to the equations for free GnRH \eqref{eq:F} and LH \eqref{eq:L} to incorporate diffusive effects. 

\section{ABBREVIATIONS}
Gonadotropin Releasing Hormone (GnRH); \sep GnRH Receptor (GnRH-R); \sep Luteinizing Hormone (LH); \sep Follicle Stimulating Hormone (FSH); \sep Mitogen-Activated Protein Kinases (MAPK); \sep MAPK phosphatases (MKP)
\section{ACKNOWLEDGMENTS}
This work was supported by grants DMS-0616710 and EF-1038593 from the National Science Foundation (MCR,HFN) and D12AP00001 (J Harer) from DARPA.
\section{HIGHLIGHTS}
\begin{itemize}
\item We created a mathematical model for GnRH induced LH secretion in perifusion cultures.
\item The model fits well to three experimental data sets.
\item Negative feedback in the signaling pathway explains the triphasic LH response.
\item Receptor desensitization explains LH secretion termination in prolonged GnRH input.
\item The pattern of LH output depends on the medium velocity and biochemical parameters.
\end{itemize}

\section{REFERENCES}
\bibliographystyle{plain}
\bibliography{paperref}
\end{document}